Title: **Modeling Suggests Combined-Drug Treatments for Disorders Impairing Synaptic Plasticity *via* Shared Signaling Pathways**

Paul Smolen[1], Marcelo A. Wood[2], Douglas A. Baxter[1], and John H. Byrne[1]

[1]Department of Neurobiology and Anatomy
W. M. Keck Center for the Neurobiology of Learning and Memory
McGovern Medical School of the University of Texas Health Science Center at Houston
Houston, Texas 77030

[2]Department of Neurobiology and Behavior
University of California Irvine
Irvine, California 92697

Running Title: Simulated Combined-Drug Treatments for Cognitive Disorders

Correspondence Address:
Paul D. Smolen
Department of Neurobiology and Anatomy
W.M. Keck Center for the Neurobiology of Learning and Memory
McGovern Medical School of the University of Texas Health Science Center at Houston
Houston, TX 77030
E-mail: Paul.D.Smolen@uth.tmc.edu
Voice: (713) 500-5601
FAX: (713) 500-0623

## Abstract

Genetic disorders such as Rubinstein-Taybi syndrome (RTS) and Coffin-Lowry syndrome (CLS) cause lifelong cognitive disability, including deficits in learning and memory. Can pharmacological therapies be suggested that improve learning and memory in these disorders? To address this question, we simulated drug effects within a computational model describing induction of late long-term potentiation (L-LTP). Biochemical pathways impaired in these and other disorders converge on a common target, histone acetylation by acetyltransferases such as CREB binding protein (CBP), which facilitates gene induction necessary for L-LTP. We focused on four drug classes: tropomyosin receptor kinase B (TrkB) agonists, cAMP phosphodiesterase inhibitors, histone deacetylase inhibitors, and ampakines. Simulations suggested each drug type alone may rescue deficits in L-LTP. A potential disadvantage, however, was the necessity of simulating strong drug effects (high doses), which could produce adverse side effects. Thus, we investigated the effects of six drug pairs among the four classes described above. These combination treatments normalized impaired L-LTP with substantially smaller individual drug 'doses'. In addition three of these combinations, a TrkB agonist paired with an ampakine and a cAMP phosphodiesterase inhibitor paired with a TrkB agonist or an ampakine, exhibited strong synergism in L-LTP rescue. Therefore, we suggest these drug combinations are promising candidates for further empirical studies in animal models of genetic disorders that impair histone acetylation, L-LTP, and learning.

**Key Words**: learning; simulation; feedback; drug combination; computational; synergism

## Introduction

Determination of gene mutations associated with disorders that cause cognitive disability is providing new opportunities for developing therapeutic interventions, including pharmacologically targeting molecules in the signaling pathway associated with the mutation (Ehninger and Silva 2011; Guilding et al. 2007; McBride et al. 2005). For example, the underlying cause of Rubinstein-Taybi syndrome (RTS) is a mutation in the lysine acetyltransferase CREB binding protein (CBP) (or less commonly a related acetyltransferase, p300) (Petrij et al. 1995; Roelfselma and Peters 2007; Rubinstein and Taybi 1963). CBP is an obligatory cofactor in the activation of transcription by cyclic AMP response element binding protein (CREB) (Roelfsema and Peters 2007). Phosphorylated CREB forms a complex with CBP, followed by induction of target genes (Chrivia et al. 1993; De Cesare et al. 1999; McManus and Henzel 2001), and CBP is critical for the formation of LTM (Alarcon et al 2004; Korzus et al. 2004; Wood et al. 2005). Animal studies suggest that some aspects of RTS-related cognitive disability can be mitigated by enhancing signaling through the cyclic AMP (cAMP) pathway (Alarcon et al. 2004; Bourtchouladze et al. 2003).

Other examples of cognitive disability are related to deficits in histone acetylation pathways. For example, Coffin-Lowry syndrome (CLS), a significant cause of human cognitive disability, is caused by mutations in the ribosomal S6 kinase 2 (RSK2) gene (Delaunoy et al. 2006). RSK2 phosphorylates CREB after activation by ERK (Rawashdeh et al. 2016; Xing et al. 1996, 1998), suggesting that in CLS, learning impairment may be due in part to deficient CBP-mediated histone acetylation. Histone acetylation mechanisms also are known to work together with nucleosome remodeling mechanisms to ultimately regulate gene expression. Similar to disorders caused by mutations in genes involved in histone acetylation, mutations in genes involved in nucleosome remodeling also give rise to cognitive disability. For example, the neuron-specific nBAF complex is composed of at least 15 subunits encoded by 28 genes (Ronan et al. 2013; Staahl and Crabtree 2013). Mutations in nBAF genes give rise to human disorders including Coffin-Siris syndrome, Nicolaides-Baraitser syndrome, and Autism Spectrum Disorder (reviewed in Vogel-Ciernia and Wood, 2014). One key nBAF neuron-specific subunit, BAF53b, has been shown to be necessary for L-LTP and several forms of learning (Vogel-Ciernia et al. 2013; White et al. 2016; Yoo et al. 2017). BAF53b forms a complex with CBP and associated proteins (Aizawa et al. 2004; Wu et al. 2007) and this interaction activates gene expression (Qiu and Ghosh 2008), suggesting BAF53b mutation can also be modeled as impairing CBP-dependent histone acetylation and gene induction. In support of this suggestion, focal deletion of histone deacetylase 3 ameliorates impairments in BAF53b mutant mice (Shu et al. 2018). Several other genetic disorders also cause cognitive disability and correlate with impaired histone acetylation, including Kabuki syndrome, for which phenotypic rescue by a histone deacetylase inhibitor is observed in a mouse model (Bjornsson et al. 2014); mutations in the TRRAP protein, which recruits acetyltransferases to chromatin (Cogne et al. 2019); and mutations in the chromatin regulator BRPF1 (Yan et al. 2017).

Given the complexities found in histone acetylation regulated processes, it is likely that concurrent targeting of multiple sites can yield substantial advantages. Combination therapies can have the advantage of drug synergism, allowing for effectiveness with substantially lower dosages of the individual drugs (Barrera et al. 2005; Zimmerman et al. 2007). It is becoming apparent that combining computational simulations of intracellular signaling pathways with *in vitro* or *in vivo* studies can greatly aid in predicting effective drug combinations (Boran and Iyengar 2010; Severyn et al. 2011; Smolen et al. 2014). Combination drug treatments that target disorders causing deficient learning and memory have not, however, yet been developed.

To address this issue, we have built a computational model of components of the intracellular signaling pathways necessary for late long-term potentiation (L-LTP), in order to examine ways in which simulations, and analysis of synergy, may help identify candidate combination pharmacotherapies for disorders causing cognitive disability. L-LTP requires gene expression and protein synthesis and is a neuronal correlate of learning and memory. Convergence of multiple kinases and the transcription factors they modify, in order to activate genes necessary for L-LTP and consequent long-term memory (LTM), suggests that simulated drug effects may multiply together in a way that generates synergistic enhancement of LTM. Development of effective therapies to compensate for this lesion is important, because impaired histone acetylation is characteristic of several disorders that lead to severe learning impairment. Therefore, we simulate effects of a specific molecular lesion – impaired histone acetylation and consequently impaired gene induction. These simulations suggests combination drug therapies that can rescue deficits, due to impaired histone acetylation, in L-LTP and LTM are candidate treatments for a considerable range of genetic disorders that cause substantial or severe deficits in learning and memory.

We explored whether our previous published model describing histone acetylation (Smolen et al. 2014); extended to represent dynamics of the growth factor denoted brain-derived neurotrophic factor (BDNF), RSK, and CaM kinase IV; could: **a)** suggest modulation of biochemical parameters as potential targets to rescue deficits in synaptic plasticity, in particular L-LTP, associated with impaired histone acetylation, **b)** identify pairs of parameters that are plausible drug targets for modulating the dynamics of signaling pathways and that, when concurrently varied, rescue the simulated deficit in L-LTP, and **c)** predict regions of drug synergism associated with concurrent adjustments to these parameter pairs. To evaluate drug synergism, two complementary methods were used: additive synergism and nonlinear blending.

### Measures of Synergism

Synergism occurs when drugs reinforce each other so that their combined effect exceeds the prediction from their separate effects (Bijnsdorp et al. 2011; Zhang et al. 2014). A straightforward measure, applicable in any situation, is additive synergism, also denoted as response additivity or a linear interaction effect (Foucquier and Guedj 2015; Slinker 1998). By this measure, with fixed doses of drugs 1 and 2, the combined response to 1 and 2 combined exceeds the summed responses to 1 alone and to 2 alone. Additive synergism, if present, implies that synergism also exists according to a second commonly used measure, Bliss independence (Bliss 1939; Berenbaum 1989). This implication follows because when calculating the Bliss independence measure, the product of the individual effects of the drugs (expressed as fractions less than 1) is subtracted from the summed responses to the individual drugs. A second measure, nonlinear blending, has been argued to have significant advantages in that drug interactions such as coalism and potentiation do not decrease its usefulness (Peterson and Novick 2007; Zhang et al. 2014). To assess nonlinear blending (NB) synergism, individual doses of drugs 1 and 2 are first fixed so that each give comparable, not saturated, responses (using a common response variable). Then the response is quantified for a series of drug mixtures. One varies the proportion of drug 1 from 0 to 100% of its effective individual dose while concurrently varying the amount of drug 2 oppositely, from 100% to 0 of its effective dose. The resulting dose-response curve will be concave down if there is NB synergism. If the curve has its maximum at an end point, then weak NB synergism is present, but may not be useful because the maximal response is obtained at one endpoint, using only one drug. However, strong NB synergism can also occur, with the maximum response away from the end points at a mixture of drugs 1 and 2. This is an important advantage, because then a drug mixture clearly gives a superior effect compared to either individual drug (Foucquier and Guedj 2015).

Figure 1 illustrates additive and NB synergism. The upper curve is concave down with a maximum away from the edges, illustrating strong NB synergism. The lower curve is a summed effect dose-response curve. This curve illustrates the response magnitude that would be predicted for a given mixture by simple addition of the separately simulated, single-drug responses. If the NB curve, which gives the actual response, lies above the summed effect curve's prediction, additive synergism occurs.

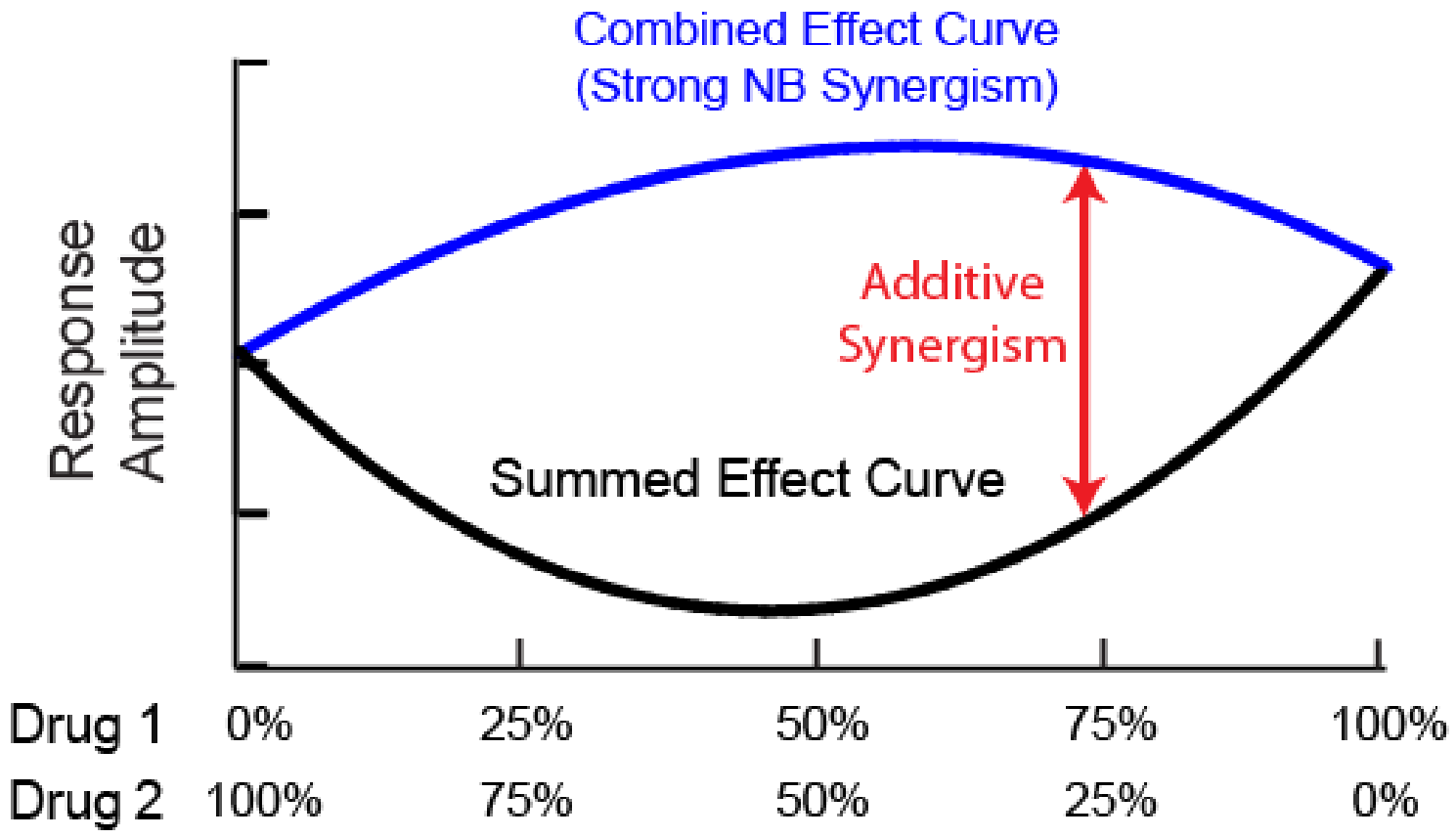

FIGURE 1. **Additive synergism and strong nonlinear blending synergism.** The effective dose of Drug 1 increases from left to right while that of Drug 2 concurrently decreases. The two endpoints illustrate similar single-drug response amplitudes for each drug. The lower black curve illustrates the predicted effect of the drug combinations, with each point obtained by adding the separate effects of drugs 1 and 2 at a given percent mixture. The upper curve illustrates the actual effect of the drug combinations.

**Model Development**

The present model is an elaboration of a previous model by Smolen et al. (2014). The present model incorporates several key processes that are essential for a more complete understanding of mechanisms of LTP. Figure 2 schematizes the model with new pathways denoted by orange arrows. Stimuli, such as a tetanus, are represented by simultaneous brief elevations (for 1 s) of synaptic $Ca^{2+}$, nuclear $Ca^{2+}$, and cAMP. The rate of activation of Raf, upstream of ERK, is also briefly elevated (for 1 min). Stimulus parameters are as in Smolen et al. (2014). This stimulus activates CaMKII, PKA, and ERK (Fig. 2). In turn, CaMKII, ERK, and PKA each phosphorylate a component of a synaptic tag. This tag is required for synaptic "capture" of proteins necessary for late LTP (Barco et al. 2002; Frey and Morris 1997). The degrees of phosphorylation of these three sites are denoted Tag-1 through Tag-3. The amount of active synaptic tag is denoted by a variable TAG and is set equal to the product of Tag-1, Tag-2, and Tag-3. Parameters for the kinetics of the tag and the kinases that set the tag are as in Smolen et al. (2014).

Empirically, during the formation of LTM, PKA can phosphorylate CREB at Ser 133, activating transcription (Matsushita et al. 2001). ERK can phosphorylate several transcription factors (TFs) including AP-1 (Guedea et al. 2011) and Elk1 (Cammarota et al. 2000). In the model, PKA and ERK each phosphorylate a site on a TF or co-activator, and the PKA TF is denoted CREB. In addition, the ribosomal S6 kinase p90 isoform (RSK2), encoded by *rsk2*, activates CREB after activation by ERK (Rawashdeh et al. 2016; Xing et al. 1996, 1998). In *Aplysia*, inhibition of RSK expression or activity reduces long-term synaptic facilitation, suggesting RSK is required for learning-related synaptic plasticity (Liu et al. 2020). Thus in the model, RSK and PKA both phosphorylate the CREB site, in a simply additive manner. CaMKIV phosphorylates CBP on Ser 301, and mutating this site impairs NMDA-stimulated transcription (Impey et al. 2002). CaMKIV knockout mice are deficient in contextual fear LTM (Song et al. 2015). Thus in the model, CaMKIV phosphorylates a third TF site. The degrees of phosphorylation of these TF sites are denoted by the variables PCREB, PERK, and PCK4 (Methods, Eqs. 27-29).

PCREB, PERK, and PCK4 act multiplicatively to drive sequential histone acetylations (or analogous sequential chromatin modifications) necessary for gene induction. In the model two sequential acetylations are assumed. The states with 0 – 2 acetylations are denoted AC0 – AC2. AC2 induces transcription and consequent synthesis of a plasticity-related protein (PRP) essential for L-LTP. We use CCAAT enhancer binding protein (C/EBP) as an exemplar PRP because: 1) expression of the β isoform of C/EBP is necessary for consolidation of inhibitory avoidance (IA) learning (Taubenfeld et al. 2001), plausibly corresponding to late hippocampal LTP, and 2) the expression of C/EBPβ is activated by CREB (Niehof et al. 1997; Zhang et al. 2004).

Recent data demonstrate that BDNF signaling is essential for several forms of LTP and learning, and therefore it is necessary to incorporate BDNF signaling into the model. For example, structural LTP of excitatory hippocampal synapses requires BDNF (Colgan et al. 2018) and cortical BDNF is required for motor learning (Andreska et al. 2020). As with C/EBP, histone acetylation is assumed to drive increased synthesis of BDNF. Empirically, CREB is required for *bdnf* induction (Tao et al. 1998). In the model, C/EBP as well as CREB are required for BDNF synthesis, motivated by recent data describing BDNF expression during IA learning (Bambah-Mukku et al. 2014). In turn, BDNF feeds back to further activate the Raf -> ERK pathway. In keeping with observations that do not show long-lasting ERK activation after inhibitory avoidance learning (Chen et al. 2011), this feedback is set to be relatively weak. However, this positive feedback is sufficient to substantially prolong the transient ERK activation after LTP induction, enhancing L-LTP. To simplify the representation of BDNF synthesis, release, and action, the model represents two homogenous pools of BDNF, intracellular and extracellular. The respective concentrations are denoted by variables BDNF(int) and BDNF(ext). BDNF(int) is synthesized in response to increased histone acetylation (AC2), assuming the same equations and parameter values as for C/EBP synthesis. BDNF(int) is released, increasing (BDNF)ext, abruptly in response to an LTP-inducing stimulus. In turn, BDNF(ext) activates the ERK cascade by linearly increasing the rate constant for Raf activation. Small basal rate constants for BDNF and C/EBP synthesis and BDNF release are assumed. BDNF and C/EBP degrade with first-order kinetics.

In the model, CaMKII, PKA, and ERK converge (orange ellipse) to induce synthesis of the kinase denoted protein kinase M ζ (PKMζ). Empirically, inhibition of CaMKII, PKA, or ERK impairs PKMζ translation (Kelly et al. 2007). PKMζ translation is increased by LTP induction (Hernandez et al. 2003). PKMζ appears necessary for L-LTP, because induction of tetanic L-LTP is blocked by dominant negative PKMζ (Ling et al. 2002). Persistent activation of PKMζ is also necessary for maintenance of L-LTP and memory (Hsieh et al. 2017; Ling et al. 2002; Pastalkova et al. 2006). Thus, our model assumes PKMζ activity is necessary for L-LTP. We do not model late maintenance or reactivation over days or longer, therefore PKMζ activity is not persistent (i.e., not bistable). However, the time constant for PKMζ deactivation is relatively slow, ~ 8 h. We note that other studies used an inducible (Volk et al. 2013) or a constitutive (Lee et al. 2013) knockout of PKMζ to argue that PKMζ is not necessary for L-LTP. However, compensation by other protein kinase C isoforms appears to occur with a constitutive knockout (Tsokas et al. 2016), and a study with a different inducible knockout found impairment of L-LTP and LTM (Wang et al. 2016). We also note Rossato et al. (2019) suggest PKMζ may be necessary for reconsolidation of established LTM following reactivation, rather than maintenance of LTM during periods without reactivation. However, overall, we believe there is still substantial evidence that PKMζ is necessary for maintenance of at least some forms of late LTP and LTM, perhaps during reconsolidation.

Increases in the amounts of the synaptic tag (TAG), C/EBP, and PKMζ converge, with these variables multiplying together to yield the rate of increase of a synaptic weight W. The resulting increase in W corresponds to L-LTP. A slow time constant for decay of W corresponds to persistence of L-LTP for several days in the absence of additional stimuli. As compared to Smolen et al. (2014), the new pathways and equations in this model are: 1) activation of CaMKIV and consequent phosphorylation of a TF site necessary for transcription of PRPs, 2) synthesis of PKMζ and its necessity to increase W, 3) activation of RSK by ERK and phosphorylation of CREB by RSK, and 4) synthesis and release of BDNF, activating Ras/Raf. Model equations, variables, and parameter values are given in Methods.

Except for the BDNF positive feedback loop described above, the structure of the model consists of unidirectional, parallel signaling cascades from stimulus input, through protein synthesis and synaptic tagging, and to synaptic weight. Given this structure, the only potential for multiple stable solutions for a given input strength is if the BDNF feedback loop is strong enough to induce bistability. For the control parameter set (normal induction of L-LTP) and for the parameter sets corresponding to rescue of impaired L-LTP by an ampakine or a cAMP phosphodiesterase inhibitor (Table I) we verified the feedback loop is too weak to sustain bistability, for normal and 10 x normal stimulus parameters. Bistability was also not found during other simulations of drug effects. Thus the model, with control parameter values, does not have degenerate solutions.

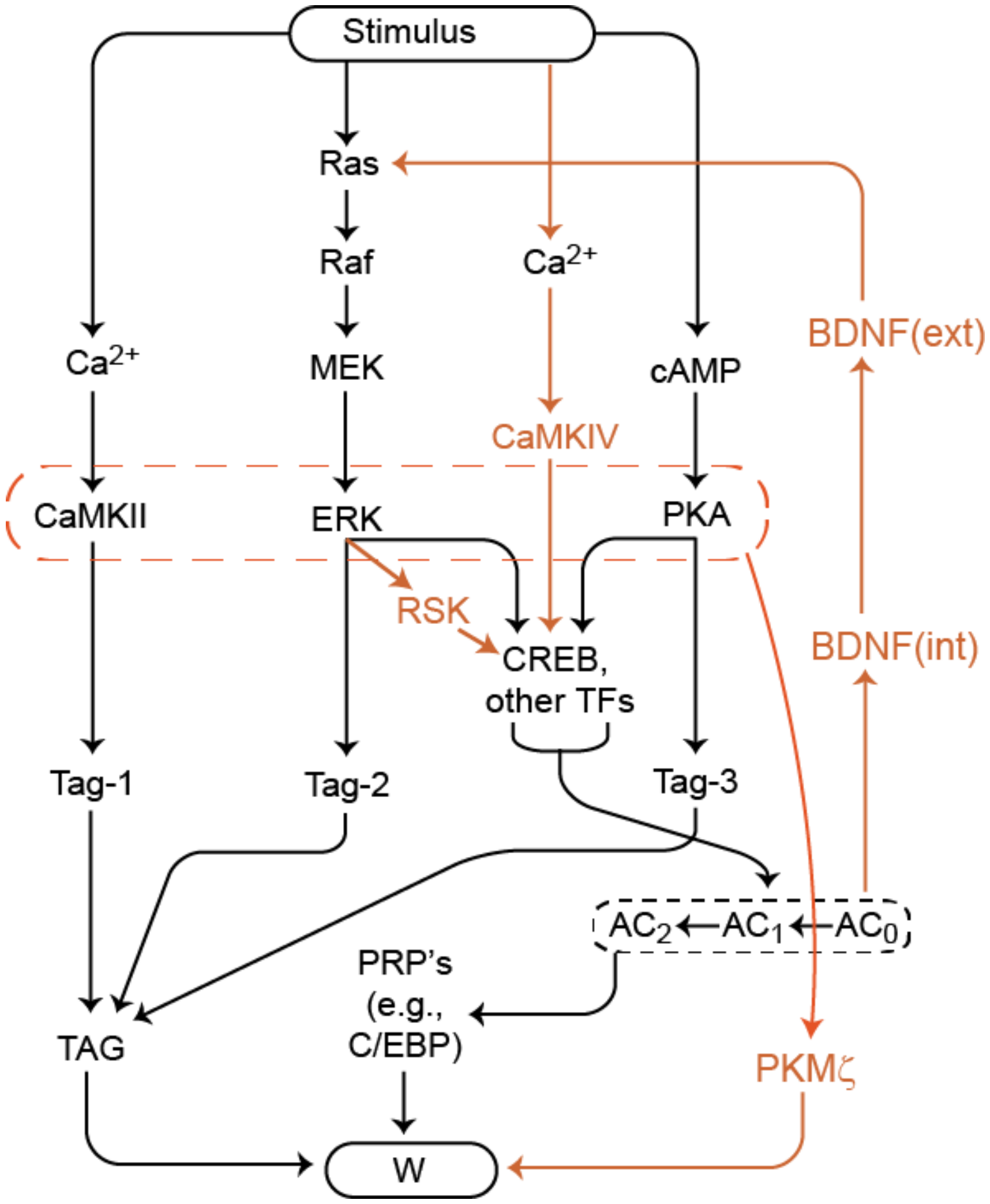

FIGURE 2. **Model describing dynamics of intracellular signaling pathways essential for the induction of late LTP.** Stimuli are represented by simultaneous brief elevations of synaptic $Ca^{2+}$, nuclear $Ca^{2+}$, and cAMP. The rate of activation of Raf, upstream of ERK, is also briefly elevated. These elevations activate CaMKII, CaMKIV, PKA, and ERK→RSK. CaMKII, ERK, and PKA each phosphorylate a site of a synaptic tag. New pathways in this model as compared to Smolen et al. (2014) are denoted by orange arrows, and ellipse. CaMKII, PKA, and ERK converge (orange ellipse) to induce synthesis of PKMζ. PKA, ERK, and CaMKIV each phosphorylate a site on a transcription factor (TF) or co-activator, with PKA and RSK additively phosphorylating a CREB site. These phosphorylations drive sequential histone acetylations necessary for gene induction. Gene induction is represented by increase of a plasticity-related protein, C/EBP, and by induction of BDNF, which feeds back to activate the Raf -> ERK pathway. Synthesized intracellular BDNF is released upon LTP induction, elevating extracellular BDNF, which activates the Ras→ERK pathway. Increased TAG, C/EBP, and PKMζ converge to increase a synaptic weight W. A slow time constant for W yields L-LTP persistence for several days.

## Results

### Simulation of LTP

In addition to simulating the dynamics of CaM kinases II and IV, PKA, synaptic and nuclear ERK activation as in Smolen et al (2014), the expanded model describes simulated dynamics of released BDNF (BDNFext), PKMζ, and RSK activation (Figs. 3A1-A2), following a stimulus of three brief tetanic current injections, with an inter-stimulus interval (ISI) of 5 min. Similarly spaced tetani commonly induce L-LTP (Frey and Morris 1997, Huang and Kandel 1995). Each tetanus was modeled with brief, concurrent increases in synaptic and nuclear $Ca^{2+}$ concentrations, cAMP concentration, and first-order rate constant for Raf activation (Smolen et al. 2006). Simulated dynamics of CaM kinases, ERK, and PKA (Fig. 3A1) are qualitatively similar to empirical data (Bhalla and Iyengar 1999; Komiyama et al. 2002; Murakoshi et al. 2017; Roberson and Sweatt 1996; Rosenblum et al. 2002; Wu et al. 2001), as is the magnitude and time course of the increase in synaptic weight W (Fig. 3A3). The positive feedback from BDNF synthesis to Raf → ERK activation prolongs ERK activation, increasing its decay time by ~ 1 h with respect to no feedback (not shown), but this feedback strength is not close to enabling bistability. RSK activity exhibits a large fold increase from basal to peak (Fig. 3A2). PKMζ activity increases relatively rapidly, over the initial ~20 min after the onset of stimulus (Fig. 3A2). Due to the model's assumed dependence of PKMζ activation on CaMKII activity, the rapid increase in PKMζ activity is primarily driven by the fast, high-amplitude increases in CaMKII activity caused by the tetani. Because CaMKII does not remain active for more than several min post-tetanus in this model, PKMζ activity does not continue to increase. Instead, following stimulus, PKMζ activity declines towards a low basal level over several hours, consistent with its assumed slow deactivation time constant. Because we do not model early LTP, the slow rise in W is similar to empirical L-LTP induction by BDNF, bypassing early LTP (Ying et al. 2002). Also similar to data (Frey and Morris 1997, 1998), the synaptic tag required for L-LTP (variable TAG) returns to baseline after 2-3 h (Fig. 3A3). After TAG returns to baseline W can no longer increase (Eq. 37), corresponding to an inability of the synapse to capture plasticity-related proteins. The increase in the plasticity-related protein C/EBP overlaps with the TAG increase (Fig. 3A3). The magnitude of L-LTP (per cent increase in W) is 142%.

We further examined the model dynamics for two perturbations of critical parameters. Figure 3B illustrates the effect of substantially increasing the basal activation of the Raf→ERK pathway, by increasing the rate constant for Raf activation ($k_{fbasRaf}$, Methods, Eq. 9). The activation of CaM kinases and PKA is not affected, but ERK activation is elevated, by similar amounts for both peak and basal ERK (Fig. 3B1). Basal and peak PKMζ activity is increased substantially, due to activation of PKMζ by ERK. RSK activation, downstream of ERK, is also substantially increased, as is histone acetylation (AC2), which is driven indirectly by PKA, ERK, RSK, and CaMKIV (Fig. 3B2, note change in scale). These kinase activities converge to enhance expression of the synaptic tag (TAG) and C/EBP, which in turn further converge, greatly enhancing the elevation of synaptic weight W (Fig. 3B3, note change in scale). This simulation illustrates the nonlinear convergence of signaling pathways.

Figure 3C illustrates the effect of strengthening the BDNF positive feedback loop by increasing the rate constant coupling BDNF elevation to Raf activation ($k_{bdnf}$, Methods, Eq. 9). Although the increase is substantial, enhanced positive feedback only further prolongs the activation of ERK (Fig. 3C1, note change in time scale) and RSK (Fig. 3C2). Bistability does not occur (i.e., there is no permanent transition of ERK activity to a high steady state). The lack of bistability is due to the dependence of BDNF synthesis on other kinases (CaMKIV, PKA) which are only activated by the stimulus, and are required to increase histone acetylation. Nonetheless, the increased and prolonged ERK activation does result in enhanced synaptic tagging and C/EBP synthesis (Fig. 3C3), and their convergence in turn greatly enhances the elevation of synaptic weight.

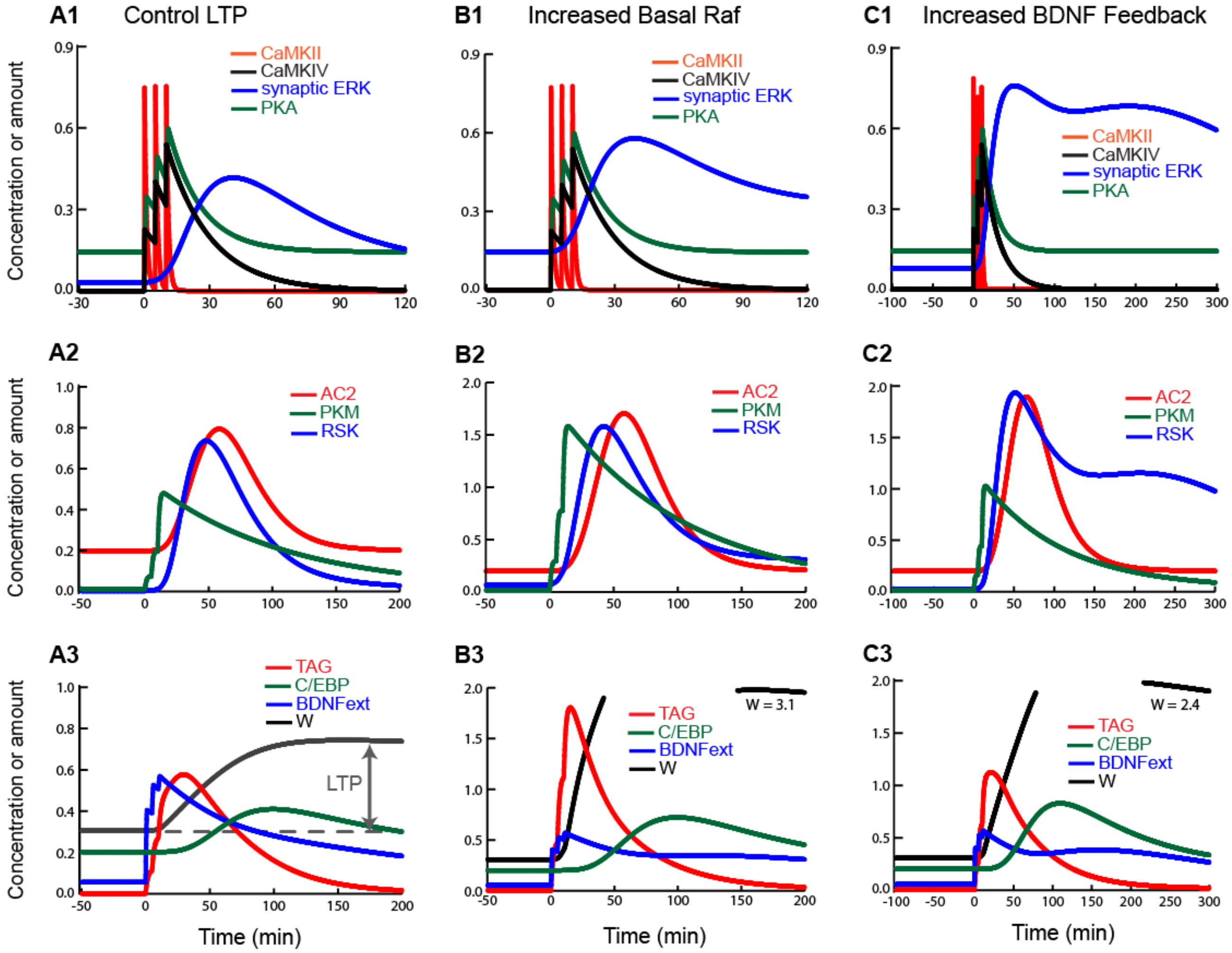

FIGURE 3. **Simulation of normal LTP, increased basal Raf activation, and increased positive feedback. (A1 – A3)** Normal LTP. Control values for all parameters are used (Methods). **(A1)** Time courses of kinase

activities. CaMKII tracks the spaced tetanic stimuli. **(A2 – A3)** Time courses of histone acetylation (AC2, the most acetylated state), synaptic ERK activity, PKMζ activity, the synaptic tag, extracellular BDNF, C/EBP, and W. L-LTP is assessed as the per cent increase in W 3 h after the onset of tetanic stimuli. Note the difference in time scales between A1 and A2 – A3. Plotted concentrations are arbitrary units because time courses are vertically scaled for ease of visualization. Scaling factors for Figs. 3-5 are: CaMKII 0.1, CaMKIV 1.0, PKA 40.0, synaptic ERK 4.5, RSK 65.0, TAG 250.0, AC2 130.0, BDNFext 7.0, C/EBP 0.6, PKM 25.0, W 0.1. **(B1 – B3)** Enhanced ERK activation, increased downstream C/EBP synthesis and synaptic tagging, and enhanced LTP due to an increase in the rate constant for basal Raf activation, $k_{fbasraf}$, from 0.006 $min^{-1}$ to 0.01 $min^{-1}$. **(C1 – C3)** Enhanced and prolonged ERK activation following an increase in the strength of BDNF positive feedback, and increased C/EBP synthesis, synaptic tagging, and LTP. The rate constant $k_{bdnf}$ is increased from 0.04 $min^{-1}$ to 0.15 $min^{-1}$.

## Parameter sensitivity

To determine the sensitivity of the model to parameter changes, the dependence on the amount of L-LTP to relatively minor changes (20%) of all parameter values was analyzed, perturbing from the 76 control values listed in Methods (8 stimulus amplitudes and durations, 68 other model parameters). We used a standard analysis defining a set of relative sensitivities $S_i$, with i ranging over all parameters $p_i$ (Frank, 1978). R denotes the magnitude of stimulus response (i.e.,L-LTP). For each $p_i$, a change is made, and the resulting change in R is determined. Relative sensitivity $S_i$ is defined as the absolute value of the <u>relative,</u> or <u>fractional,</u> change in R divided by the <u>relative</u> change in $p_i$,

$$S_i = \left| \frac{\Delta R / R}{\Delta p_i / p_i} \right|$$

L-LTP was quantified 3 h post-stimulus. Sensitivity to most of the 76 parameter variations, including all stimulus amplitudes and durations, was not large, with $S_i$ below 3. There were eight exceptions with $S_i$ between 3.0 and 10.0. Interestingly, all but two of these exceptions were for Raf-ERK pathway parameters (Methods, Eqs. 8-18). These exceptions were: 20% increases in the total concentration of Raf kinase ($[Raf_{tot}]$), in the rate constant for Raf activation ($k_{fbasRaf}$), in the rate constants for MEK activation ($k_{fMEK}$) and inactivation ($k_{bMEK}$), in the inactivation rate constant for Raf ($k_{bRaf}$), and in the ERK activation rate constant ($k_{fERK}$). The remaining two exceptions were the basal level of cAMP ($cAMP_{bas}$) and the basal rate constant for histone acetylation ($k_{facbas}$). This lack of unduly large sensitivity for most parameter variations (68 out of 76) suggests the model is relatively robust and can be regarded as a reasonable qualitative model of signaling cascades important for L-LTP. Further modeling studies might seek to add buffering mechanisms, especially within the Raf-ERK pathway, to reduce sensitivity to these specific parameters.

## Simulation of impaired LTP

To model the impairment in L-LTP associated with reduced lysine acetyltransferase activity in RTS, the rate constants driving histone acetylation were reduced (Smolen et al. 2014). Figure 4B, when compared with Fig. 4A, illustrates profound impairment of L-LTP when the rate constants for stimulus-induced histone acetylation ($k_{fac}$, Eq. 30) and basal (unstimulated) histone acetylation ($k_{facbas}$, Eq. 30) are reduced, plausibly corresponding to a heterozygous mutation of CBP (as in RTS) or of associated proteins such as BAF53b (Introduction). Reduced histone acetylation strongly impairs synthesis of BDNF and C/EBP. The activation of upstream kinases is not affected, except that ERK and RSK activation is lessened due to reduced positive feedback from impaired BDNF synthesis. Synaptic tagging, which does not depend on transcription, is also not affected, but induction of C/EBP and BDNF is greatly impaired (Fig. 4A3). Late LTP is reduced by about two thirds (to 49%), similar to the reduction in a mouse model of RTS (Alarcon et al. 2004).

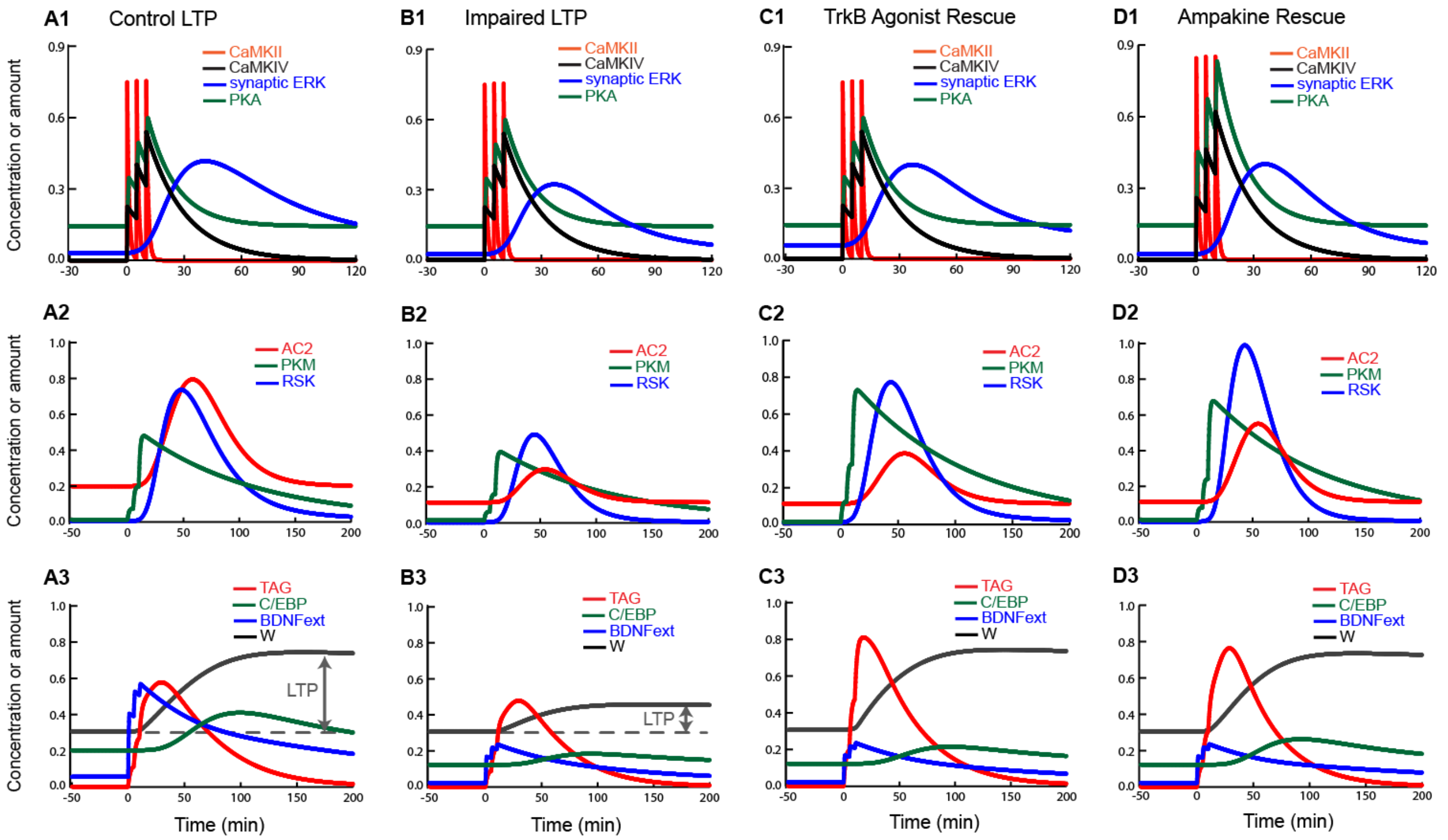

FIGURE 4. **Simulation of impairment of L-LTP by reduced histone acetylation, and rescue of impaired L-LTP by two drug types. (B1 – B3)** Normal LTP as in Fig. 3A **(B1 – B3)** Impairment of C/EBP and BDNF synthesis, and L-LTP, when histone acetylation rate constants are decreased. The rate constant for stimulus-induced acetylation, $k_{fac}$, is decreased from 12.0 $min^{-1}$ to 7.0 $min^{-1}$, and that for basal acetylation, $k_{facbas}$, is decreased from 0.004 $min^{-1}$ to 0.003 $min^{-1}$. L-LTP is greatly reduced, as are the increases in acetylation, BDNF, and C/EBP. **(C1 – C3)** Rescue by a TrkB agonist, which persistently upregulates basal Raf activation, which "primes" the ERK -> RSK pathway for stimulus-induced activation and increases basal ERK activity. TAG, which depends on ERK, is enhanced. These changes compensate for impaired induction of C/EBP and BDNF, and also enhance PKMζ activation. Thus L-LTP is near normal. **(D1 – D3)** Rescue by an ampakine, which enhances activation of CaMKII, Raf -> ERK -> RSK, CaMKIV, and PKA. This enhancement increases TAG and PKMζ activation, and TF phosphorylation which augments histone acetylation and C/EBP and BDNF synthesis. L-LTP is near normal.

## Simulated rescue of LTP impairment by clinically relevant drug classes

Histone (lysine) deacetylase inhibitors (HDACI) and cAMP phosphodiesterase inhibitors (PDEI) have improved learning and memory in rodent models of RTS (Alarcon et al. 2004; Bourtchouladze et al. 2003; Kazantsev and Thompson 2008). We previously simulated rescue of deficit L-LTP in RTS with combined PDEI and HDACI treatment (Smolen et al. 2014). Recent data suggest additional drug classes may ameliorate deficits in synaptic plasticity, in RTS and other disorders that impair histone acetylation. The TrkB receptor, when activated by BDNF, activates the Ras -> ERK pathway, inducing gene expression and promoting late LTP (Messaoudi et al. 2002; Panja and Bramham 2014), thus TrkB agonists are candidates for rescuing impairments in synaptic plasticity and learning (Zeng et al. 2013). We therefore simulated the efficacy of a TrkB agonist in rescuing deficit L-LTP due to histone acetylation impairment. The effect of a TrkB agonist is expected to be constitutive, upregulating the Ras -> ERK pathway irrespective of stimuli. Thus to simulate the effect of a TrkB agonist, we increased the basal rate constant describing the activation of Raf by Ras in the absence of stimulus (parameter $k_{fbasRaf}$) (Eq. 9). This "priming" of the Raf -> ERK pathway elevates basal kinase activities and facilitates a greater stimulus response.

Figure 4C illustrates rescue of L-LTP induction, impaired as in Fig. 4B, by a TrkB agonist (TrkBAg). $k_{fbasRaf}$ is increased from its control value of 0.006 min$^{-1}$ to 0.0077 min$^{-1}$. Basal ERK activity, and the amplitude and duration of stimulus-induced ERK activation, are therefore enhanced (Fig. 4C1-C2, compared to control in Fig. 4A). The amplitude and duration of the synaptic tag are thereby enhanced (Fig. 4C3) as is PKMζ activation, which depends on ERK (Fig. 4C2). The large increases of TAG and of PKMζ multiply to augment the increase in synaptic weight W (Eq. 37) thus L-LTP is augmented to near normal (Fig. 4C3, Table 1).

Ampakines are another promising drug class. These drugs act on the AMPA receptor (AMPAR), to inhibit deactivation and/or desensitization (Arai and Kessler 2007; Lynch et al. 2014), enhancing and prolonging postsynaptic depolarization after stimulus. Ampakines improve spatial long-term memory in rats (Staubli et al. 1994), augment classical conditioning (Shors et al. 1995), and improve learning in aged rats (Lauterborn et al. 2016). We therefore simulated the efficacy of an ampakine in rescuing deficit

L-LTP, alone or combined with a TrkB agonist, HDACI, or PDEI. The enhancement of stimulus-induced depolarization due to an ampakine is expected to augment depolarization-sensitive postsynaptic influxes of $Ca^{2+}$ through NMDA receptors and voltage-dependent $Ca^{2+}$ channels, although this has not been empirically assessed. Such an increase of $Ca^{2+}$ influx would enhance activation of $Ca^{2+}$-dependent adenylyl cyclases important for LTP (Wong et al. 1999), and increase $Ca^{2+}$-dependent activation of the Ras -> ERK pathway (Schmitt et al. 2005). We therefore simulated the effect of an ampakine by increasing the amplitudes of stimulus-induced elevations of synaptic and nuclear $Ca^{2+}$, (parameters $A_{Casyn}$ and $A_{Canuc}$), and of cAMP elevation ($A_{cAMP}$), and of the increase in activation of Raf by Ras ($A_{STIM}$). In the absence of data, we assumed a common multiplying factor, $f_{AMPK}$, for these parameters, typically $f_{AMPK}$ is ~1.2. These parameter changes augment activation of CaMKII, CaMKIV, PKA, and Raf. We note that increased $Ca^{2+}$ influx will have additional effects on biochemical pathways not included in our model.

Figure 4D illustrates simulated rescue of L-LTP by an ampakine (Ampak). Augmentation of all kinase activation peaks including PKMζ occurs, compared to control. Here $f_{AMPK} = 1.25$, a 25% increase from control (1.0). As with a TrkB agonist, most of the increase in L-LTP to normal levels (Table 1) is due to ampakine-mediated augmentation, even above control values, of PKMζ and TAG. These augmentations of PKMζ levels, for both drugs, could plausibly be tested empirically.

We examined simulations of different "doses" of TrkBAg (different values of $k_{fbasRaf}$) and of Ampak (different values of $f_{AMPK}$) for trends in how the time courses of variable depend on dose. All trends were predictable, and monotonic with dose. As TrkBAg increases, both basal and peak levels of synaptic ERK increase, while activation of CaMKII, PKA, and CAMKIV is not affected. Levels of active nuclear ERK also increase (not shown), thus peak active RSK increases. Peak PKMζ increases substantially. Some of this PKMζ increase is due to the elevation of basal ERK prior to stimulus. TrkBAg fails to restore normal levels of histone acetylation (compare AC2, Fig. 4C2 vs. A2) and thus does not restore C/EBP induction or normal BDNF levels, both of which depend on acetylation and resulting gene induction (with induction only modeled as a protein synthesis rate term). However, to compensate, the peak of TAG is well above normal (due to elevated ERK) (Fig. 4C3 vs. A3) and peak PKM is above normal. For the illustrated TrkBAg dose, these changes qualitatively average out to yield approximately normal LTP (increase in W, Fig. 4C3 vs. A3).

As Ampak increases, the model assumes all signaling pathways are activated more due to increased stimulus-dependent $Ca^{2+}$ influx. Thus peak PKA, CaMKII, CaMKIV, and synaptic ERK activities monotonically increase with dose. Because PKA and synaptic ERK are both increased, the import of active nuclear ERK is cooperatively enhanced (not shown), yielding a large increase in active RSK, well above normal (Fig. 4D2 vs. A2). Increased RSK and PKA drive more TF phosphorylation and histone acetylation than with TrkBAg, although AC2 is still below normal (Fig. 4D2 vs. A2). Thus, C/EBP is somewhat higher than with TrkBAg, although still below normal. BDNF levels show little variation. Upstream, kinase activation by Ampak increases the peaks of TAG and of PKMζ. However, because basal ERK is not elevated (Ampak is modeled as only affecting stimulus response), the increases in TAG and PKM are somewhat less than with TrkBAg. To compensate, the increase in C/EBP is somewhat greater than with TrkBAg. For the illustrated Ampak effect, these changes qualitatively average out to yield approximately normal LTP.

We also simulated rescue of L-LTP induction (Table 1) by a PDE inhibitor (PDEI) and an HDAC inhibitor (HDACI), similarly to Smolen et al. (2014). The effect of a PDEI was simulated by increasing the duration of stimulus-induced cAMP elevation (the parameter $d_{CAMP}$). An increase from 1 min to 2.4 min restored normal L-LTP. The effect of HDACI was simulated by decreasing the histone deacetylation rate constant (parameter $k_{bac}$, Eqs. 32-33). A decrease from 0.1 $min^{-1}$ to 0.074 $min^{-1}$ sufficed to restore normal L-LTP amplitude. For PDEI especially, the necessary parameter change appears too great to correspond to a physiologically plausible drug effect. A motivation to simulate the drug combinations was to reduce the magnitude of these single-drug parameter changes necessary to rescue L-LTP.

**LTP rescue with reduced drug doses, by means of drug combinations**

For each of the six drug pairs corresponding to the four drug classes HDACI, PDEI, Ampak, and TrkBAg, we identified combined parameter changes that rescued L-LTP. For all combinations, the parameter change amplitudes (legend, Table 1) were substantially reduced relative to single drugs. However, the parameter changes required for rescue with PDEI remained large. None of these drug combinations significantly altered the basal, unstimulated synaptic weight. Thus, in the context of the model, only synaptic changes associated with learning are affected, not basal homeostasis.

| Condition Simulated | Late LTP (percent increase in synaptic weight) | Basal synaptic weight |
|---|---|---|
| Normal Tetanic LTP | 142 | 0.31 |
| Impaired LTP (decreased histone acetylation) | 49 | 0.31 |
| LTP Rescued by Ampak | 140 | 0.31 |
| LTP Rescued by PDEI | 141 | 0.31 |
| LTP Rescued by TrkBAg | 142 | 0.31 |
| LTP Rescued by HDACI | 137 | 0.31 |
| LTP Rescued by Ampak + PDEI | 137 | 0.31 |
| LTP Rescued by Ampak + TrkBAg | 138 | 0.31 |
| LTP Rescued by PDEI + TrkBAg | 136 | 0.31 |
| LTP Rescued by HDACI + Ampak | 128 | 0.31 |
| LTP Rescued by HDACI + PDEI | 135 | 0.31 |
| LTP Rescued by HDACI + TrkBAg | 126 | 0.31 |

TABLE 1. **Rescue of L-LTP by simulated single drugs and drug pairs.** L-LTP is assessed 2 h post-onset of three tetanic stimuli, and basal synaptic weight W is assessed prior to stimuli. For all rescues the percent amplitude of L-LTP is within 8 percent of normal, and the basal value of W (scaled as in Fig. 3) is within 1 percent of normal. Single drug doses (parameter changes) are greater than doses in drug pairs, in order to achieve rescue. For drug pairs, the changed parameter values are: TrkBAg + Ampak, $k_{fbasRaf} = 0.0068$ min$^{-1}$, $f_{amp} = 1.12$. PDEI + Ampak, $d_{CAMP} = 1.55$ min, $f_{amp} = 1.11$. HDACI + Ampak, $k_{bac} = 0.087$ min$^{-1}$, $f_{amp} = 1.125$. For TrkBAg + PDEI, $k_{fbasRaf} = 0.00675$ min$^{-1}$, $d_{CAMP} = 1.6$ min. HDACI + PDEI, $k_{bac} = 0.087$ min$^{-1}$, $d_{CAMP} = 1.7$ min. HDACI + TrkBAg, $k_{bac} = 0.087$ min$^{-1}$, $k_{fbasRaf} = 0.00685$ min$^{-1}$.

For four of the rescues by drug pairs, time courses of variables following L-LTP induction are illustrated in Fig. 5. The differences in dynamics between rescues, and between rescues and normal L-LTP (Fig. 3A), were found to be predictable in terms of drug actions. For example, in Figs. 5A, C, and D, PDEI elevates the PKA activation peak above normal. This enhances nuclear import of active ERK, and thus tends to increase RSK activation. However, the increase of RSK activity is only substantial in Fig. 5C, because for this drug pair PDEI and Ampak cooperate to further activate PKA.

In Fig. 5D, HDACI directly enhances histone acetylation. Therefore, C/EBP and BDNF levels are higher with this combination than for any other rescue, due to enhanced gene induction. For all rescues, however, C/EBP and BDNF remain below normal L-LTP levels (Fig. 3A). For the other model variables, differences in dynamics are not substantial between drug pairs.

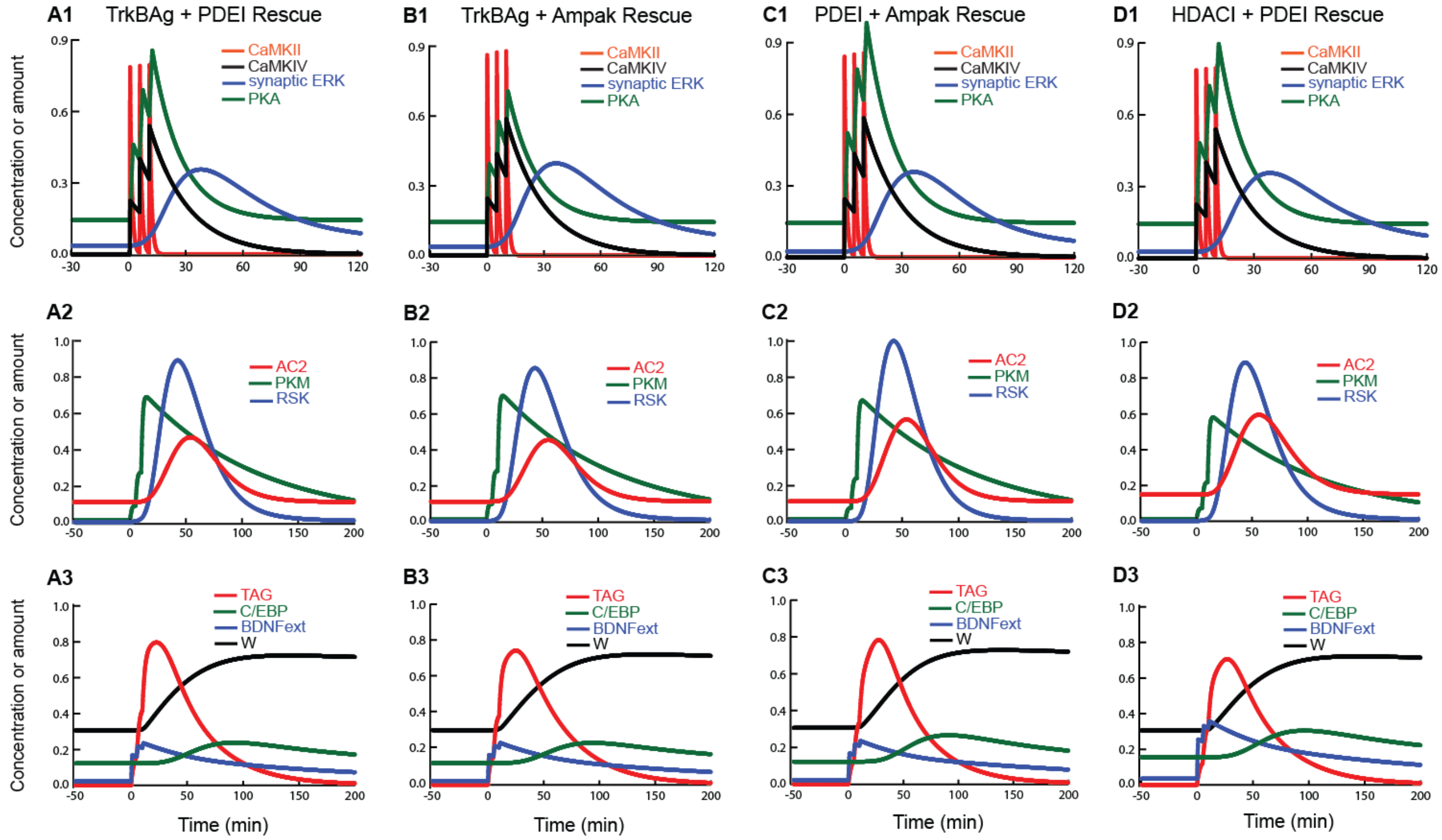

FIGURE 5. **Simulation of rescue of impaired L-LTP by four drug combinations.** Vertical scaling factors for all variables are as in Fig. 3. Parameter changes for each rescue are given in the legend for Table 1.

## Drug combinations exhibit synergistic interaction

Additive synergism was first assessed for the six drug pairs. For all pairs, synergism was evident (Fig. 6A-F). L-LTP produced by each drug pair exceeded the predicted L-LTP from the summed effects of the single drugs at the same simulated doses. Among the three drug pairs with a histone deacetylase inhibitor (Fig. 6D-F), the largest excess, and greatest L-LTP with substantial synergism, occurred for HDACI + PDEI. This result is similar to the rescue of L-LTP we found previously for this pair, with our earlier model that lacked BDNF, RSK and PKMζ (Smolen et al. 2014). For these HDACI pairs, the greatest synergism occurred near the midpoints of the illustrated dose-response curves, which correspond to half the single-drug doses (i.e., changes in $k_{bac}$, $f_{amp}$, $k_{fbasraf}$, and $d_{camp}$ relative to control were 50% of those required for single-drug rescue). As with our earlier model, HDACI + PDEI gave nearly complete rescue at the midpoint. NB synergism was then assessed for the three HDACI pairs as illustrated in Fig. 1, linearly increasing the dose of drug 1 from zero to maximum as shown while concurrently decreasing the dose of drug 2 from maximum to zero. NB synergism was not present for any of them. For HDACI +

PDEI (Fig. 6D), this result contrasts with our earlier model, which did simulate strong NB synergism (Smolen et al. 2014).

Interestingly, we did find strong NB synergism, together with strong additive synergism, in the remaining drug pairs (Fig. 6A-C). At the curve midpoints, Ampak + TrkBAg, PDEI + TrkBAg, and PDEI + Ampak yielded close to complete rescues of L-LTP (similar to corresponding values in Table 1), and the greatest additive synergism was attained near the midpoints. Furthermore, for these pairs, the simulated single-drug doses at the midpoints (changes in $f_{amp}$, $k_{fbasraf}$, and $d_{camp}$ relative to control) were less than 50% of those for single-drug rescue. With strong NB synergism the largest LTP enhancement, near the midpoints, is well above both endpoints (both single-drug effects) despite these reduced doses. The simulated effectiveness, and synergism, of these three pairs (Fig. 6A-C) appears stronger than for the three pairs that include HDACI. Considering also that strong additive synergism also implies synergism by another common measure, Bliss independence, the simulations of Fig. 6A-C support the suggestion that these pairs are candidates for further empirical testing, in *in vitro* or *in vivo* models of impaired synaptic plasticity and learning in RTS, CLS, or related disorders that impair histone acetylation and gene induction.

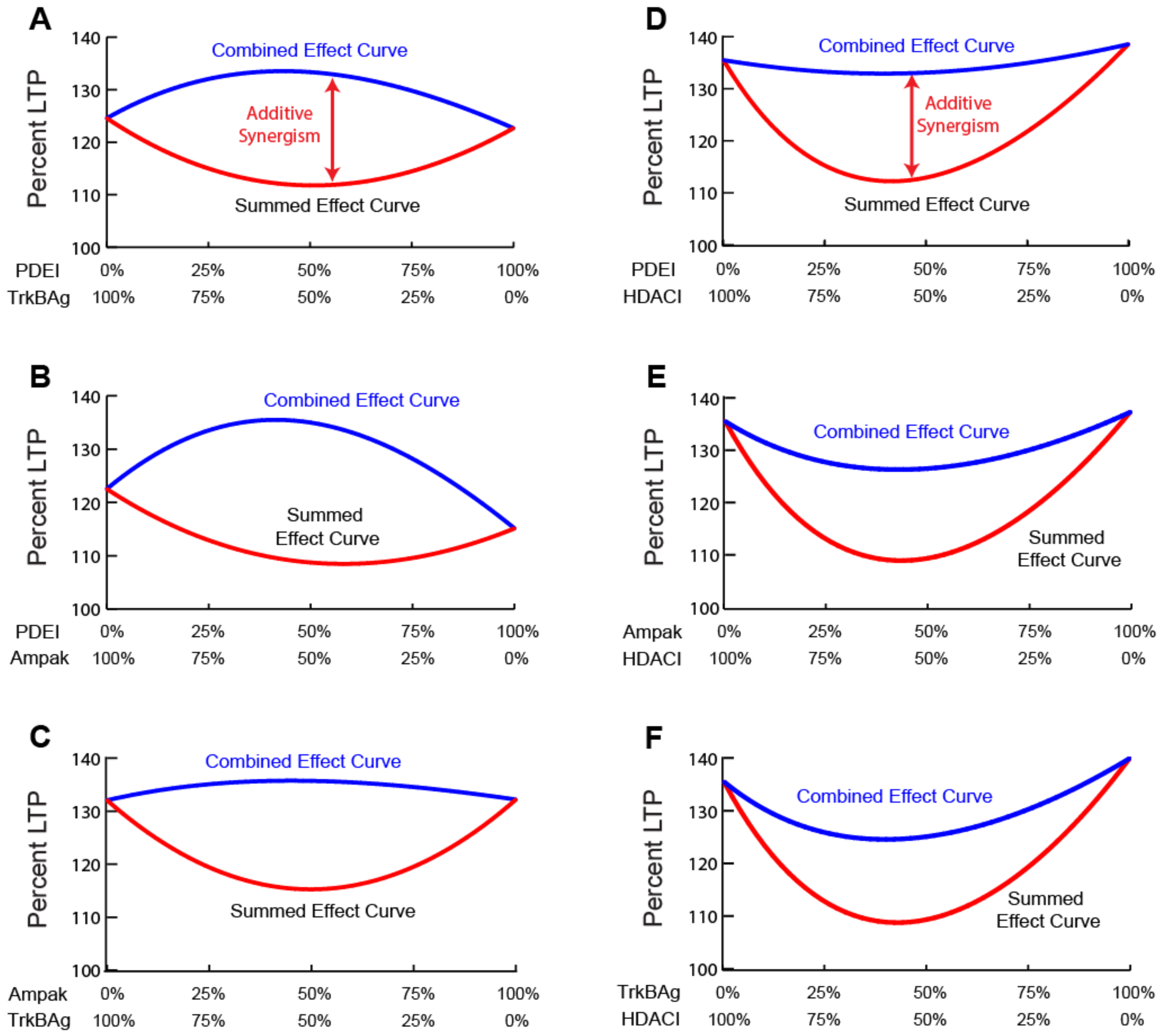

FIGURE 6. **Additive synergism and nonlinear blending synergism for rescue of L-LTP. (A)** Rescue by PDEI paired with TrkBAg. From left to right, the dose of PDEI increases from zero to maximum, while TrkBAg decreases from maximum to zero. In **(A)**, in addition to displaying strong additive synergism, the combined effect curve is concave down, with a maximum at an intermediate point, thus evidencing strong NB synergism. Strong additive synergism is also evident for **(B)** and **(C)**. **(B)** Rescue by PDEI and Ampak. (PDEI increases left to right). Strong NB synergism is evident. **(C)** Rescue by TrkBAg and Ampak. Only weak NB synergism is evident. **(D)** Rescue by PDEI and HDACI. For **(D) – (F)**, additive synergism is evident, but no NB synergism occurs (the combined-effect curves are concave down). **(E)** Rescue by Ampak and HDACI. **(F)** Rescue by TrkBAg and HDACI. Parameter variations are as follows. For **(A) – (C)**: for PDEI, $d_{camp}$ varies from 1.0 min (no drug) to 2.2 min (max drug, **A**) or 2.1 min (**C**), for TrkBAg, $k_{fbasRaf}$ varies from 0.006 min$^{-1}$ (no drug) to 0.0075 min$^{-1}$ (max drug) (**A**) or 0.0076 min$^{-1}$ (**C**), for Ampak, $f_{amp}$ varies from 1.0 (no drug) to 1.22 (max drug) (**B**) or 1.24 (**C**). For **(D) – (F)**: for HDACI, $k_{bac}$ varies from 0.1 min$^{-1}$ (no drug) to 0.0074 min$^{-1}$ (max drug), for PDEI, $d_{camp}$ varies from 1.0 min (no drug) to 2.4 min (max drug), for TrkBAg, $k_{fbasRaf}$ varies from 0.006 min$^{-1}$ (no drug) to 0.0077 min$^{-1}$ (max drug), for Ampak, $f_{amp}$ varies from 1.0 (no drug) to 1.25 (max drug).

What elements of the model determine which drug pairs yield the greatest synergism? Because the model equations are complex and often nonlinear, it does not appear possible to derive analytic expressions for the extent of synergism displayed by these drug pairs. At present, we are only able to make a qualitative observation that drugs which affect a greater number of signaling pathways tend to favor more additive and NB synergism. In the model, HDAC inhibitors act downstream of all kinase pathways and only affect synthesis of BDNF and C/EBP. In Fig. 6, the plots with HDACI (Fig. 6D-F) fail to show NB synergism, and tend to show less additive synergism than the other plots. Ampakines, by contrast, upregulate every signaling pathway in the model. Two plots with Ampak (Fig. 6B, C) show strong NB synergism and strong additive synergism. Among all pairs, the greatest amount of both forms of synergism is seen in Fig. 6B. TrkB agonists upregulate the ERK pathway, which in turn enhances both synaptic tagging and transcription, thus TrkBAg engages more pathways than HDACI, but fewer than Ampak. Two plots with TrkBAg (Fig. 6A, C) show strong NB synergism and strong additive synergism. Figure 6C illustrates a partial exception to this qualitative trend. Although it pairs the two drugs that engage the most signaling pathways (TrkBAg and Ampak), and shows both NB and additive synergism, it displays somewhat less synergism than Fig. 6B.

## Discussion

We found that a model of key biochemical pathways required for the induction of LTP could simulate impaired LTP seen in rodent models of RTS and suggest modulation of specific biochemical parameters as potential targets to rescue deficits in synaptic plasticity and learning in several disorders, including RTS and CLS, that impair CREB phosphorylation and consequent histone acetylation and gene induction. The model is a simplification of the complex processes underlying memory and the effects of impaired acetylation, and modeling of drug effects is also highly simplified. For example, increased $Ca^{2+}$ influx during stimulus, due to ampakine, can be expected to affect the dynamics of numerous enzymes dependent on $Ca^{2+}$-calmodulin, including phosphatases and numerous kinases, and $Ca^{2+}$ binding to annexins affects vesicle transport and membrane organization. Despite these simplifications, we believe the model, and particularly the simulations of Fig. 6, suggest drug combinations that could be tested in animal models for rescue of LTP deficits.

Specifically, for a TrkB agonist paired with a PDE inhibitor, a TrkB agonist with an ampakine, and a PDE inhibitor paired with an ampakine, concurrent parameter variations were simulated for which additive synergism was maximized and LTP is simultaneously restored to near-normal values. These parameter variations correspond to the ranges on the combined effect curves that are the greatest distance above the summed effect curves. For these three pairs, strong NB synergism was also simulated, with LTP rescue by drug combinations being substantially greater than the corresponding single-drug effects at higher doses. Subsequent empirical studies might initially develop dose-effect relationships that allow translation of parameter changes in these ranges to drug doses.

A limitation of using models such as ours to predict effective combined pharmacotherapies will be the incompleteness of models describing LTP induction combined with empirical uncertainties in parameter values. In model construction, parameter values are generally not all available from a single experimental preparation, or a single species of animal. Instead, data from several types of studies (*e.g.,* slice and cell culture) and animals (*e.g.* rodents and primates) need to be used (Bhalla and Iyengar 1999; Hayer and Bhalla 2005; Smolen et al. 2006, 2012). It has also been shown that different induction protocols engage, in part, different molecular signaling pathways. For example, hippocampal L-LTP induced by four 100 Hz tetani appears CBP-independent, whereas L-LTP induced by a single 100 Hz tetanus paired with a dopamine D1 agonist is CBP-dependent (Wood et al. 2005), and CREB-dependent (Pittenger et al. 2002). However, theta-burst L-LTP requires CREB, BDNF, and BAF53b, suggesting this stimulus engages histone acetylation and nucleosome remodeling to induce genes required for L-LTP (Patterson et al. 2001; Vogel-Ciernia et al. 2013; White et al. 2016). Allowing for these significant caveats, we believe that the potential benefits of this strategy that uses modeling to suggest drug combinations for further empirical studies are substantial.

**Empirical data support the pharmaceutical efficacy of small molecules in the classes simulated here**

Recent data have identified a promising TrkB small molecule agonist, 7,8-dihydroxyflavone (7,8-DHF). This compound ameliorates spatial memory deficits in a mouse Alzheimer's disease (AD) model (Gao et al. 2016), and cognitive and dendritic spine abnormalities in a mouse Fragile X model (Tian et al. 2015). Additional small molecule TrkB agonists have also been considered as possible pharmacotherapies (Zeng et al. 2010).

Considerable recent data have supported the efficacy of ampakines. Two ampakines, CX-929 and CX-1846, have been shown to rescue deficits in LTP and spatial learning due to ageing (Lauterborn et al. 2016; Radin et al. 2016). CX-929 also rescues deficits in LTP and in fear conditioning in a mouse model of Angelman syndrome (Baudry et al. 2012). CX-691 has been reported to rescue a spatial learning deficit in a rat AD model (Mozafari et al. 2018). Treatment with CX-516 rescues deficits in LTP and in fear conditioning in another mouse model, of intellectual disability due to TM4SF2 deletion (Murru et al. 2017). The availability of tested TrkB agonists and ampakines, which cross the blood-brain barrier in rodents, strongly suggests testing of the efficacy of this drug combination in animal models of genetic disorders such as CLS that correlate with impaired histone acetylation, as we discuss previously, as well as in animal models of the other disorders above, such as Fragile X and AD. Recently 7,8-DHF and CX-929 have separately been reported to rescue deficient spatial learning in a mouse Fragile X model, with 7,8-DHF also rescuing deficient LTP (Seese et al. 2019).

The particularly strong synergism, as measured additively and with nonlinear blending, for the ampakine – TrkB agonist pair (Fig. 6C) suggests that with this pair, substantially greater rescue of LTP and plausibly learning, will be observed when the drugs are paired together, at similar effective individual doses, then when either is given separately at a larger dose. In addition, the amplitude of parameter changes required for LTP rescue are smaller for this pair than for all others (Table 1). The rate constant $k_{fbasRaf}$, for constitutive Raf activation, was increased by 11%, from 0.006 min-1 to 0.0068 min-1, and the factor $f_{amp}$, for stimulus enhancement by an ampakine, was increased by 12%, from 1 to 1.12. Changes of this magnitude appear to correspond to physiologically plausible drug effects.

With regard to cAMP PDE inhibitors, both rolipram and a newer compound, HT-0712, have been shown to rescue deficits in long-term memory in a mouse RTS model (Bourtchouladze et al. 2003). HT-0712 also enhances contextual and trace conditioning in aged mice, and enhances induction of gene expression by CREB in these mice (Peters et al. 2014).

Although our simulations did not identify NB synergism between HDAC inhibitors and either PDE inhibitors, TrKB agonists, or ampakines, additive synergism was identified in all cases. Empirically combining voronistat, a HDACI, with tadalafil, a PDEI, has rescued plasticity and learning deficits in a mouse model of Alzheimer's disease (Cuadrado-Tejedor et al. 2015). Induction of numerous genes has been observed upon induction of late LTP (Chen et al. 2017). Therefore we believe it is of interest to empirically investigate drug combinations that include HDACIs to ameliorate learning impairments in cognitive disorders. HDAC inhibition alone has been shown capable of transforming a transient form of early LTP (independent of CREB, CBP, and transcription) into L-LTP that requires transcription (Vecsey et al. 2007), an effect our model may not capture due to its lack of representation of early LTP. The empirical rescue in Cuadrado-Tejedor et al. (2015) occurred at physiologically tolerated doses. These data, and our simulated additive synergism, suggest combinations of HDACIs and PDEIs remain candidates for testing in animal models of genetic disorders with impaired histone acetylation.

## Methods

### Model equations, stimuli, and parameter values

We first describe stimuli that induce LTP, then describe equations representing kinase cascades, and then describe equations representing downstream components – synaptic tagging, gene induction, and synaptic weight increase. LTP is induced by three tetanic stimuli with a standard inter-stimulus interval of 5 min. As in our previous study (Smolen et al. 2006), each tetanus is modeled as inducing: 1) a square-wave rise in synaptic and nuclear $Ca^{2+}$ concentrations, $[Ca_{syn}]$ and $[Ca_{nuc}]$, to values $A_{Casyn}$ (1 μM control value) and $A_{Canuc}$ (0.5 μM control value) respectively. These $Ca^{2+}$ concentrations are otherwise at a low basal value $Ca_{bas}$ (40 nM control). The brief increase has a duration $Ca_{dur}$. $Ca_{dur}$ was chosen as 3 s for each 1-s, 100-Hz tetanus. A similar duration of $Ca^{2+}$ elevation is suggested by data, with imaging suggesting a time constant of 1–2 s for decay of $Ca^{2+}$ transients after tetanus (Pologruto et al. 2004): 2) A brief elevation of cAMP. The kinetics of cAMP production and its activation by $Ca^{2+}$ have not been well characterized. Therefore, we assumed each tetanus induced a prescribed, square-wave elevation of [cAMP]. Data suggest the time for [cAMP] to return to basal levels after stimulation is ~1–2 min (Vincent and Brusciano 2001). Thus we assumed [cAMP] remained elevated for 1 min (model parameter $d_{cAMP}$) during and after stimulation. Otherwise cAMP is at a basal value $cAMP_{bas}$.

Each tetanus also transiently activates the ERK signaling pathway. Neuronal ERK signaling can be activated by $Ca^{2+}$ elevation acting via CaM kinase I (Schmitt et al. 2005) or by cAMP elevation (Morozov et al. 2003) or by a $Ca^{2+}$-independent pathway involving mGluR5 (Yang et al. 2004). Raf activation is the convergence point for these mechanisms of ERK cascade activation. Rather than modeling these complexities in detail, we assumed each tetanus increased the zero-order rate constant for synaptic and somatic Raf phosphorylation and activation (Eq. 9). In the absence of detailed data, we assumed a square-wave increase lasting for 1 min. The zero-order rate constant has a basal value of $k_{fbasRaf}$ and is increased for 1 min to an elevated value of $(A_{STIM} + k_{fbasRaf})$, with $A_{STIM} = 0.15\ min^{-1}$.

CaMKII and CaMKIV are activated by Hill functions of cytoplasmic and nuclear $Ca^{2+}$ concentrations. CaMKIV also requires its upstream kinase, CaM kinase kinase (CaMKK), to be similarly activated by nuclear $Ca^{2+}$. We use Hill functions to describe these activations. In Eqs. 1–2, the Hill coefficients are given standard values of 4. These Hill functions constitute a minimal representation of the activation of CaM kinases by calmodulin (CaM), because four $Ca^{2+}$ ions bind cooperatively to CaM and CaM-$Ca_4$ activates CaM kinases. For CAMKII, data show a steep $[Ca^{2+}]$ dependence that can be characterized by a Hill coefficient ≥4 (Bradshaw et al. 2003). For CAMKIV activity, a similar steep $[Ca^{2+}]$ dependence is likely, given CaM-$Ca_4$'s obligatory binding to both CAMKIV and CAMKK. In Eqs. 3-5, and subsequently, the symbol $\tau$ is used to denote time constants that describe first-order deactivation or decay, e.g., $\tau_{ck2}$ describes deactivation of CaMKII.

$$f_1 = \frac{[Ca_{syn}]^4}{[Ca_{syn}]^4 + K_{Casyn}{}^4} \tag{1}$$

$$f_2 = \frac{[Ca_{nuc}]^4}{[Ca_{nuc}]^4 + K_{Canuc}{}^4} \tag{2}$$

$$\frac{d[CaMKII_{act}]}{dt} = k_{fck2}\, f_1 - \frac{[CaMKII_{act}]}{\tau_{ck2}} \tag{3}$$

$$\frac{d[CaMKK_{act}]}{dt} = k_{fckk}\, f_2 - \frac{[CaMKK_{act}]}{\tau_{ckk}} \tag{4}$$

$$\frac{d[CaMKIV_{act}]}{dt} = k_{fck4}\, f_2 - \frac{[CaMKIV_{act}]}{\tau_{ck4}} \tag{5}$$

For cAMP to activate PKA, two cAMP molecules must bind cooperatively to the regulatory subunit of a PKA holoenzyme (Herberg et al. 1996). Thus, our qualitative representation of PKA activation assumes the activation rate is a Hill function of the second power of [cAMP].

$$f_3 = \frac{[cAMP]^2}{[cAMP]^2 + K_{cAMP}{}^2} \tag{6}$$

$$\frac{d[PKA_{act}]}{dt} = \frac{(f_3 - [PKA_{act}])}{\tau_{PKA}} \tag{7}$$

Tetani are assumed to phosphorylate and activate the first kinase in the ERK cascade, commonly Raf-1 or B-Raf in neurons (Agell et al. 2002). Active Raf phosphorylates MAP kinase kinase (MAPKK) twice, activating MAPKK. MAPKK then phosphorylates ERK twice, activating ERK. Saturable Michaelis-Menten terms have been commonly used to describe phosphorylations and dephosphorylations of the kinases in the ERK cascade (e.g., Pettigrew et al. 2005), and are used here. Phosphatase dynamics are not explicitly modeled. Only doubly phosphorylated MEK and ERK are considered active. Raf can also be activated due to an increase in extracellular BDNF as described further below. These dynamics can be described by the following differential equations. Concentrations of singly phosphorylated forms are given by conservation conditions.

$$\left[\mathrm{RAF}\right] = \mathrm{RAF}_{\mathrm{TOT}} - \left[\mathrm{RAFP}\right] \qquad 8)$$

$$\frac{d\left[\mathrm{RAFP}\right]}{dt} = \left(\mathrm{A}_{\mathrm{STIM}} + k_{\mathrm{fbasRaf}}\right)\left[\mathrm{RAF}\right] + k_{\mathrm{bdnf}}\left[\mathrm{BDNF(ext)}\right]\left[\mathrm{RAF}\right] - k_{\mathrm{bRaf}}\left[\mathrm{RAFP}\right] \qquad 9)$$

$$\frac{d\left[\mathrm{MEK}\right]}{dt} = -k_{\mathrm{fMEK}}\left[\mathrm{RAFP}\right]\frac{\left[\mathrm{MEK}\right]}{\left[\mathrm{MEK}\right]+K_{\mathrm{MEK}}} + k_{\mathrm{bMEK}}\frac{\left[\mathrm{MEKP}\right]}{\left[\mathrm{MEKP}\right]+K_{\mathrm{MEK}}} \qquad 10)$$

$$\left[\mathrm{MEKP}\right] = \mathrm{MEK}_{\mathrm{TOT}} - \left[\mathrm{MEK}\right] - \left[\mathrm{MEKPP}\right] \qquad 11)$$

$$\frac{d\left[\mathrm{MEKPP}\right]}{dt} = k_{\mathrm{fMEK}}\left[\mathrm{RAFP}\right]\frac{\left[\mathrm{MEKP}\right]}{\left[\mathrm{MEKP}\right]+K_{\mathrm{MEK}}} - k_{\mathrm{bMEK}}\frac{\left[\mathrm{MEKPP}\right]}{\left[\mathrm{MEKPP}\right]+K_{\mathrm{MEK}}} \qquad 12)$$

$$\frac{d\left[\mathrm{ERK}\right]}{dt} = -k_{\mathrm{fERK}}\left[\mathrm{MEKPP}\right]\frac{\left[\mathrm{ERK}\right]}{\left[\mathrm{ERK}\right]+K_{\mathrm{ERK}}} + k_{\mathrm{bERK}}\frac{\left[\mathrm{ERKP}\right]}{\left[\mathrm{ERKP}\right]+K_{\mathrm{ERK}}} \qquad 13)$$

$$\left[\mathrm{ERKP}\right] = \mathrm{ERK}_{\mathrm{TOT}} - \left[\mathrm{ERK}\right] - \left[\mathrm{ERKPP}\right] \qquad 14)$$

$$\frac{d\left[\mathrm{ERKPP}\right]}{dt} = k_{\mathrm{fERK}}\left[\mathrm{MEKPP}\right]\frac{\left[\mathrm{ERKP}\right]}{\left[\mathrm{ERKP}\right]+K_{\mathrm{ERK}}} - k_{\mathrm{bERK}}\frac{\left[\mathrm{ERKPP}\right]}{\left[\mathrm{ERKPP}\right]+K_{\mathrm{ERK}}} \qquad 15)$$

The model has two compartments, synaptic and somatic, with separate ERK cascades in each compartment. Exchange between compartments is not explicitly modeled. The dynamics of PKA and of extracellular BDNF are assumed the same for both compartments. The somatic RAF -> MEK -> ERK cascade allows phosphorylation of a transcription factor site to induce *c/ebp* transcription. For Eqs. 8-14, the somatic and synaptic forms coincide. In the somatic analogue of Eq. 15, there is an additional term describing import of somatic activated ERK to the nucleus, given below. The rate constants, Michaelis constants, and total kinase concentrations are assumed identical for the two cascades. Thus we only distinguish the somatic cascade by the variant of Eq. 15, by a modified conservation condition for total somatic ERK, by an additional differential equation for imported nuclear ERK, and by labeling the somatic variables and stimulus amplitude. These somatic variables and the distinct equations are as follows:

$A_{STIMSOM}$, $RAF_{SOM}$, $RAFP_{SOM}$, $MEK_{SOM}$, $MEKP_{SOM}$, $MEKPP_{SOM}$, $ERK_{SOM}$, $ERKP_{SOM}$, $ERKPP_{SOM}$, $ERKPP_{NUC}$

$$\frac{d[\mathrm{ERKPP_{SOM}}]}{dt} = k_{fERK}[\mathrm{MEKPP_{SOM}}]\frac{[\mathrm{ERKP_{SOM}}]}{[\mathrm{ERKP_{SOM}}]+K_{ERK}} - k_{bERK}\frac{[\mathrm{ERKPP_{SOM}}]}{[\mathrm{ERKPP_{SOM}}]+K_{ERK}} - k_{nucin}[\mathrm{PKA_{act}}][\mathrm{ERKPP_{SOM}}] + k_{nucout}[\mathrm{ERKPP_{NUC}}] \quad 16)$$

$$[\mathrm{ERKP_{SOM}}] = \mathrm{ERK_{TOT}} - [\mathrm{ERK_{SOM}}] - [\mathrm{ERKPP_{SOM}}] - [\mathrm{ERKPP_{NUC}}] \quad 17)$$

$$\frac{d[\mathrm{ERKPP_{NUC}}]}{dt} = k_{nucin}[\mathrm{PKA_{act}}][\mathrm{ERKPP_{SOM}}] - k_{nucout}[\mathrm{ERKPP_{NUC}}] \quad 18)$$

Active nuclear ERK phosphorylates and activates RSK. The number of phosphorylations required to activate RSK is not certain, but from several potential phosphorylation sites, two are established as targets of ERK (Ser363 and Thr573) (Anium and Blenis 2008). Therefore, we used the same equation form for RSK activation as for ERK activation by dual phosphorylation, and assumed both phosphorylations are necessary for RSK activation.

$$\frac{d[\mathrm{RSK}]}{dt} = -k_{fRSK}[\mathrm{ERKPP_{NUC}}]\frac{[\mathrm{RSK}]}{[\mathrm{RSK}]+K_{RSK}} + k_{bRSK}\frac{[\mathrm{RSKP}]}{[\mathrm{RSKP}]+K_{RSK}} \quad 19)$$

$$[\mathrm{RSKP}] = \mathrm{RSK_{TOT}} - [\mathrm{RSK}] - [\mathrm{RSKPP}] \quad 20)$$

$$\frac{d[\mathrm{RSKPP}]}{dt} = k_{fRSK}[\mathrm{ERKPP_{NUC}}]\frac{[\mathrm{RSKP}]}{[\mathrm{RSKP}]+K_{RSK}} - k_{bRSK}\frac{[\mathrm{RSKPP}]}{[\mathrm{RSKPP}]+K_{RSK}} \quad 21)$$

Protein kinase M $\zeta$ (denoted PKM for brevity) is rapidly translated after stimulus and is constitutively active. Its translation is activated by CaMKII, synaptic ERK, and PKA (Kelly et al. 2007). As a minimal representation of these dynamics, the following differential equation is assumed, including a small basal translation rate.

$$\frac{d[PKM_{act}]}{dt} = k_{transpkm}[CaMKII_{act}][ERKPP][PKA_{act}] + k_{transbaspkm} - k_{dpkm}[PKM_{act}] \qquad 22)$$

As discussed in our preceding publications (Smolen et al. 2006, 2014), data suggests CaMKII, PKA, and ERK all play roles in setting the synaptic tag required for induction of late LTP. The molecular nature of the tag has not been characterized. Here, as in Smolen et al. (2014), we assume three phosphorylation sites, one for each kinase, and represent the amount of active synaptic tag (TAG) as the product of the fractional phosphorylations of each site. The differential equations for the dynamics of these fractional phosphorylations are as follows:

$$\frac{d(\text{Tag-1})}{dt} = k_{phos1}[CaMKII_{act}](1-\text{Tag-1}) - k_{deph1}\text{Tag-1} \qquad 23)$$

$$\frac{d(\text{Tag-2})}{dt} = k_{phos2}[PKA_{act}](1-\text{Tag-2}) - k_{deph2}\text{Tag-2} \qquad 24)$$

$$\frac{d(\text{Tag-3})}{dt} = k_{phos3}[ERKPP](1-\text{Tag-3}) - k_{deph3}\text{Tag-3} \qquad 25)$$

$$TAG = (\text{Tag-1})(\text{Tag-2})(\text{Tag-3}) \qquad 26)$$

Nuclear ERK, PKA, and CaMKIV each phosphorylate a transcription factor or cofactor (TF) site, that each play essential roles in the induction of *c/ebp*. The TF phosphorylated by PKA is denoted CREB, and CREB is also phosphorylated by RSK at the same site. The differential equations for the fractional phosphorylations of these sites are as follows:

$$\frac{d(PCREB)}{dt} = \{k_{phos4}[PKA_{act}] + k_{phosRSK}[RSKPP]\}(1-PCREB) - k_{deph4}PCREB \qquad 27)$$

$$\frac{d(PERK)}{dt} = k_{phos5}[ERKPP_{NUC}](1-PERK) - k_{deph5}PERK \qquad 28)$$

$$\frac{d(PCK4)}{dt} = k_{phos6}[CaMKIV_{act}](1-PCK4) - k_{deph6}PCK4 \qquad 29)$$

Two sequential histone acetylations are assumed necessary to drive *c/ebp* expression. The acetylation rate is the same for each step and depends on the degree of phosphorylation of each of the three TFs. Hill coefficients of 1 are assumed for these dependencies, given the absence of information suggesting more complex functions.

$$R_{AC} = k_{facbas} + k_{fac}\frac{PCREB}{PCREB+K_{CREB}}\frac{PERK}{PERK+K_{ERK}}\frac{PCK4}{PCK4+K_{CK4}} \quad 30)$$

Each acetylation step is reversible, and a conservation condition is used for the state with zero acetylations.

$$AC0 = AC_{TOT} - AC1 - AC2 \quad 31)$$

$$\frac{d(AC1)}{dt} = R_{AC}(AC0 - AC1) - k_{bac}(AC1 - AC2) \quad 32)$$

$$\frac{d(AC2)}{dt} = R_{AC}AC1 - k_{bac}AC2 \quad 33)$$

For Eqs. 30-33, as well as for other processes (e.g. gene induction as a result of acetylation), available data is not sufficient to determine the best forms for descriptive equations, therefore we have chosen simple forms that we believe to plausibly represent these complex kinetics.

The rate of synthesis of C/EBP protein increases proportionally to the amount of the acetylation state with two acetylations, i.e., the intermediate *c/ebp* mRNA dynamics are not modeled. C/EBP also has a small basal synthesis rate and first-order degradation.

$$\frac{d[C/EBP]}{dt} = k_{trans}AC2 + k_{transbas} - \frac{[C/EBP]}{\tau_{CEBP}} \quad 34)$$

Intracellular BDNF is likewise synthesized in response to acetylation, and its synthesis rate also increases with C/EBP concentration, as suggested by data. It can also be released to extracellular BDNF. Stimulus induces rapid release. The stimulus release rate constant, $A_{RELB}$ in Eq. 36, is 0 in the absence of stimulus and is set to 1.0 min$^{-1}$, for a duration of 1 min, at the time of each tetanus. There is a small basal release rate constant $A_{BASB}$ in stimulus absence. Both forms of BDNF have first-order degradation.

$$\frac{d[BDNF(int)]}{dt} = k_{trans2}AC2\frac{[C/EBP]}{[C/EBP]+K_{CEBP}} + k_{transbas2} - \frac{[BDNF(int)]}{\tau_{BDNF}} - (A_{RELB} + A_{BASB})[BDNF(int)] \quad 35)$$

$$\frac{d\left[\mathrm{BDNF(ext)}\right]}{dt} = \left(A_{RELB} + A_{BASB}\right)\left[\mathrm{BDNF(int)}\right] - \frac{\left[\mathrm{BDNF(ext)}\right]}{\tau_{BDNF}} \qquad 36)$$

To close the BDNF positive feedback loop, extracellular BDNF enhances Raf phosphorylation, on the right-hand side of Eq. 9 above. The intermediate steps (e.g. BDNF receptor kinetics) in Raf activation are not explicitly modeled.

The rate of increase of the synaptic weight W is given as a product of the amount of TAG, the concentration of C/EBP, the activity of PKM, and the concentration of a limiting protein $P_{lim}$ introduced (Smolen et al. 2006) to prevent excessive weight increase upon repeated stimulation. There is also a small basal rate of increase, and first-order decay.

$$\frac{dW}{dt} = k_{ltp}\mathrm{TAG}\left[\mathrm{C/EBP}\right]\frac{\left[P_{lim}\right]}{\left[P_{lim}\right]+K_{lim}} + k_{ltpbas} - \frac{W}{\tau_{ltp}} \qquad 37)$$

Because the rate of $P_{lim}$ consumption is assumed proportional to the rate of increase of W during LTP, the differential equation for $P_{lim}$ is similar to that for W.

$$\frac{d\left[P_{lim}\right]}{dt} = -k_{Pl}\mathrm{TAG}\left[\mathrm{C/EBP}\right]\frac{\left[P_{lim}\right]}{\left[P_{lim}\right]+K_{lim}} + k_{Plbas} - \frac{P_{lim}}{\tau_{Pl}} \qquad 38)$$

Control parameter values are used in all simulations except as noted otherwise. Standard stimulus amplitudes and durations are given above. The 68 remaining control values are:

$Ca_{bas}$ = 0.04 μM, $cAMP_{bas}$ = 0.03 μM, $K_{casyn}$ = 0.7 μM, $K_{canuc}$ = 0.3 μM, $k_{fck2}$ = 200 $min^{-1}$ μM, $k_{fckk}$ = 25 $min^{-1}$ μM, $k_{fck4}$ = 10 $min^{-1}$ μM, $\tau_{ck2}$ = 1 min, $\tau_{ckk}$ = 0.2 min, $\tau_{ck4}$ = 20 min, $K_{camp}$ = 0.5 μM, $\tau_{PKA}$ = 15 min.

$RAF_{TOT}$ = $MEK_{TOT}$ = $ERK_{TOT}$ = 0.25 μM, $k_{fbasRaf}$ = 0.006 $min^{-1}$, $k_{braf}$ = 0.12 $min^{-1}$, $k_{fMEK}$ = 0.6 $min^{-1}$, $k_{bMEK}$ = 0.025 $min^{-1}μM^{-1}$, $K_{MEK}$ = 0.25 μM, $k_{fERK}$ = 0.52 $min^{-1}$, $k_{bERK}$ = 0.025 $min^{-1}$ $μM^{-1}$, $K_{ERK}$ = 0.25 μM, $k_{nucin}$ = 100 $min^{-1}$ $μM^{-1}$, $k_{nucout}$ = 2.5 $min^{-1}$.

$k_{trans2}$ = 5.0 $min^{-1}$ μM, $k_{transbas2}$ = 0.0001 $min^{-1}$ μM, $\tau_{BDNF}$ = 100 min, $K_{CEBP}$ = 1.0 μM, $A_{BASB}$ = 0.001 $min^{-1}$, $k_{bdnf}$ = 0.04 $min^{-1}$ $μM^{-1}$.

$RSK_{TOT}$ = 0.25 μM, $k_{fRSK}$ = 0.5 $min^{-1}$, $k_{bRSK}$ = 0.025 $min^{-1}$, $K_{RSK}$ = 0.25 μM.

$k_{transpkm}$ = 6.5 min$^{-1}$ μM$^{-2}$, $k_{transbaspkm}$ = 0.000005 min$^{-1}$ μM, $k_{dpkm}$ = 0.01 min$^{-1}$.

$k_{phos1}$ = 0.05 min$^{-1}$, $k_{deph1}$ = 0.02 min$^{-1}$, $k_{phos2}$ = 2.0 min$^{-1}$, $k_{deph2}$ = 0.2 min$^{-1}$, $k_{phos3}$ = 0.06 min$^{-1}$, $k_{deph3}$ = 0.017 min$^{-1}$, $k_{phos4}$ = 0.06 min$^{-1}$, $k_{phosRSK}$ = 0.07 min$^{-1}$, $k_{deph4}$ = 0.03 min$^{-1}$, $k_{phos5}$ = 4.0 min$^{-1}$, $k_{deph5}$ = 0.1 min$^{-1}$, $k_{phos6}$ = 0.012 min$^{-1}$, $k_{deph6}$ = 0.03 min$^{-1}$.

$k_{fac}$ = 12.0 min$^{-1}$, $k_{facbas}$ = 0.004 min$^{-1}$, $K_{CREB}$ = 1.0 μM, $K_{ERK}$ = 1.0 μM, $K_{CK4}$ = 1.0 μM, $AC_{TOT}$ = 1.0, $k_{bac}$ = 0.1 min$^{-1}$, $k_{trans}$ = 2.0 min$^{-1}$ μM, $k_{transbas}$ = 0.0003 min$^{-1}$ μM, $\tau_{CEBP}$ = 100 min.

$k_{ltp}$ = 7000 min$^{-1}$ μM$^{-2}$, $k_{ltpbas}$ = 0.0017 min$^{-1}$, $\tau_{ltp}$ = 1800 min, $K_{lim}$ = 0.3 μM, $k_{Pl}$ = 180 min$^{-1}$ μM$^{-1}$, $k_{Plbas}$ = 0.006 min$^{-1}$ μM, $\tau_{Pl}$ = 100 min.

For some parameters, empirical data were used to determine values. As described in Smolen et al. (2006), the model's total concentrations of Raf, MEK, and ERK, and the peak concentration of active PKA as determined by stimulus and kinetic parameters, are similar to experimental values. The peak concentration of active CaMKII due to a tetanus is approximately 11% of the estimated total CaMKII concentration of 70 μM (Bhalla and Iyengar 1999). Simulated peak concentrations of active CaMKIV and CaMKK are ~5-10% of the total concentrations of these enzymes (Bhalla and Iyengar 1999). Data were also used to constrain the amplitude and duration of $Ca^{2+}$ elevations, and the duration of cAMP elevation (Smolen et al. 2006).

However, for many parameters, such as rate constants *in vivo*, data are lacking to determine values. Therefore, these parameters were chosen by trial and error to yield dynamics of variables that were similar to empirical time courses when these are available. Parameters were chosen so that the amplitude of normal and impaired LTP, the time courses of CaM kinase activities, the duration of the synaptic tag, and the time required for late LTP to be induced in the absence of early LTP (i.e. the rise time of W), are similar to data, as follows. After the tetani, CAMKII decays with a time constant of ~1 min, and CAMKIV with a time constant of ∼40 min. The simulated rapid decrease of CaMKII activity after tetanus agrees with data describing CaMKII deactivation in dendritic spines after glutamate application (Murakoshi et al. 2017). The time required for decay of CAMKIV activity is also similar to data (Wu et al. 2001). PKA activity increases by ∼150% during L-LTP induction, which is similar to data (Roberson and Sweatt 1996). Simulated synaptic and somatic ERK activity both last ∼90 min. Data concerning the duration of ERK activity are contradictory. One study suggests ERK remains phosphorylated for 8 h after tetanus (Ahmed and Frey 2005). However, other studies (English and Sweatt 1997; Liu et al. 1999) suggest a much briefer activation of ∼30 min. Because long-lasting ERK activity could regulate transcription involved in L-LTP, further experimental study of ERK kinetics is warranted. Similarl to data (Frey and Morris 1997, 1998), the synaptic tag required for L-LTP (variable TAG) returns to baseline within ~3 h. L-LTP induction nears completion in ∼2 h (Fig. 3A3, time course of W). Similarly, empirical induction of L-LTP with BDNF (bypassing early LTP) requires ∼2 h (Ying et al. 2002). In Fig. 3A3, W increases by 142%. This amplitude is similar to the excitatory postsynaptic potential increase observed after three or four 1-s, 100-Hz tetani (English and Sweatt 1997; Woo et al. 2000). Simulated C/EBP induction was required to have a similar rate of increase, and a similar duration, to data describing the induction of immediate-early genes associated with LTP, such as Arg3.1/Arc (Hevroni et al. 1998; Waltereit et al. 2001).

We note that our parameter value set cannot be claimed to be unique, and that the simulated time courses of some variables, such as histone acetylation states, are not constrained by data.

### Numerical methods

The forward Euler method integrated differential equations. For the simulations of Fig. 3A and 4C, we verified fourth order Runge-Kutta (Press et al. 2007) did not give significant differences. The time step was 10 msec. Variables were initialized to 0.0001 (μM or arbitrary units) excepting those variables given by conservation conditions. Before induction of plasticity, variables were equilibrated for at least two simulated days and the slowest variable, W, was fixed at the basal value determined by the other variables. A Java program to generate the simulations in Figs. 3A and 4B-D is given as a supplementary text file online, TextS1.txt. A Java program to generate the simulations of Fig. 6B is given as a second supplementary file, TextS2.txt. The model, with parameter values to reproduce these simulations, will be submitted to the ModelDB database, and to GitHub, upon manuscript acceptance. Programs are also available by request.

## Acknowledgements

We thank Y. Zhang for comments on an earlier draft of the manuscript. Supported by NIH grants NS102490 (to JHB) and AG051807 and AG057558 (to MAW).